\documentclass{iopart}
\usepackage{iopams}
\usepackage{amssymb}
\usepackage{bm}
\usepackage{graphicx}
\begin{document}

\title[Exponential beams of electromagnetic radiation]{Exponential beams of electromagnetic radiation\footnote{We dedicate this paper to the memory of Edwin Power who recognized in his book \cite{power} the power of Riemann-Silberstein vector although he has not been aware of its early history.}}
\author{Iwo Bialynicki-Birula}
\address{Center for Theoretical Physics, Polish Academy of Sciences,\\ Al. Lotnik\'ow 32/46, 02-668 Warsaw, Poland}
\ead{birula@cft.edu.pl}
\author{Zofia Bialynicka-Birula}
\address{Institute of Physics, Polish Academy of Sciences,\\ Al. Lotnik\'ow 32/46, 02-668 Warsaw, Poland}
\begin{abstract}
We show that in addition to well known Bessel, Hermite-Gauss, and Laguerre-Gauss beams of electromagnetic radiation, one may also construct exponential beams. These beams are characterized by a fall-off in the transverse direction described by an exponential function of $\rho$. Exponential beams, like Bessel beams, carry definite angular momentum and are periodic along the direction of propagation, but unlike Bessel beams they have a {\em finite energy} per unit beam length. The analysis of these beams is greatly simplified by an extensive use of the Riemann-Silberstein vector and the Whittaker representation of the solutions of the Maxwell equations in terms of just one complex function. The connection between the Bessel beams and the exponential beams is made explicit by constructing the exponential beams as wave packets of Bessel beams.
\end{abstract}
\vspace{2cm}

\hspace{1.7cm}
Published in J. Phys. B: At. Mol. Opt. Phys. {\bf 39}, 545 (2006).
\maketitle
\vspace{1cm}

\section{Introduction}

Directed beams are very common forms of electromagnetic radiation, especially in the optical range. Several mathematical representation of such beams were introduced in the past. Among various beams, the Bessel beams and the Laguerre-Gauss beams play a distinguished role because they are characterized by a definite value of the projection of the angular momentum on the direction of propagation. Beams carrying angular momentum were studied theoretically and experimentally (a collection of papers on the optical angular momentum has been recently published \cite{oam}). A presentation of new examples of such beams seems to be worthwhile. In this paper we produce a new family of beams --- the exponential beams. The exponential beams are exact beam-like solutions of the Maxwell equations with a definite value of angular momentum and with an exponential fall-off in the transverse direction. They can be placed halfway between the Bessel beams and the Laguerre-Gauss beams. Similarly to the Bessel beams, exponential beams have a definite value of the wave vector in the direction of propagation and like the Laguerre-Gauss beams they are confined in the transverse direction. The exponential beams are not monochromatic, but their spectrum may be highly peaked. They can be conveniently represented as wave packets of Bessel beams. In our analysis of exponential beams we shall employ the tool developed in our recent work \cite{bb1} --- the Whittaker representation of the Riemann-Silberstein (RS) vector --- that enables one to derive any solution of the free Maxwell equations from a complex solution of the scalar wave equation. We begin with a brief summary of this representation.

\section{Whittaker representation of the RS vector}

A natural tool in the analysis of the solutions of the Maxwell equations, in both classical and quantum theories, is a complex vector ${\bi F}$,
\begin{eqnarray}\label{rs}
{\bi F} = \sqrt{\frac{\epsilon_0}{2}}({\bi E}+ \rmi c{\bi B}),
\end{eqnarray}
that carries all the information about the electromagnetic field. We named this vector in \cite{pwf} the
Riemann-Silberstein (RS) vector because it made its first appearance in \cite{weber} and was more extensively analyzed in \cite{sil}. The RS vector has been occasionally used in the past \cite{bateman,stratton} in classical electrodynamics, but we believe that its power lies also in bridging a gap between the classical and the quantum theory of the electromagnetic field \cite{bb1}.

Owing to the linearity and the homogeneity of the Maxwell equations, the coefficient in the definition (\ref{rs}) is not important and we shall often disregard the question of the overall normalization of ${\bi F}$ and use the RS vector in the form ${\bi F} = {\bi E}+ \rmi c{\bi B}$. Two pairs of Maxwell equations expressed in terms of ${\bi F}$ reduce to just one pair
\begin{eqnarray}\label{max}
\partial_t{\bi F}({\bi r},t) = -\rmi c\nabla\times{\bi F}({\bi r},t),\;\;\nabla\!\cdot\!{\bi F}({\bi r},t)=0.
\end{eqnarray}
Following Whittaker \cite{whitt,whitt1}, we shall express the solutions of these equations by one complex function $\chi({\bi r},t)$ (in fact, Whittaker used an equivalent representation of the solutions of real Maxwell equations in terms of two real functions)
\begin{eqnarray}\label{solf}
\left(\begin{array}{c}
F_x \\
F_y \\
F_z \end{array}\right)
=
\left(\begin{array}{c}
\partial_x\partial_z + (\rmi/c)\partial_y\partial_t\\
\partial_y\partial_z - (\rmi/c)\partial_x\partial_t\\
-\partial_x^2-\partial_y^2\end{array}\right)\chi({\bi r},t),
\end{eqnarray}
where $\chi({\bi r},t)$ obeys the d'Alembert equation
\begin{eqnarray}\label{helm}
(\frac{1}{c^2}\partial_t^2-\Delta)\chi({\bi r},t)=0.
\end{eqnarray}
In the cylindrical coordinates $(\rho,\phi,z)$, that are useful in the description of beams carrying angular momentum, the components of the RS vector are
\begin{eqnarray}\label{solfc}
\left(\begin{array}{c}
F_\rho \\
F_\phi \\
F_z \end{array}\right)
=
\left(\begin{array}{c}
\partial_\rho\partial_z + (\rmi/c\rho)\partial_\phi\partial_t\\
(1/\rho)\partial_\phi\partial_z - (\rmi/c)\partial_\rho\partial_t\\
-\partial_\rho^2-(1/\rho)\partial_\rho-(1/\rho^2)\partial_\phi^2
\end{array}\right)\chi(\rho,\phi,z,t).
\end{eqnarray}

\section{Exponential beams}

We shall construct exponential beams as wave packets of Bessel beams. In this manner we can automatically obtain a spectral decomposition of exponential beams, since every Bessel beam is monochromatic. The Bessel beams are obtained from the Whittaker representation (\ref{solf}) or (\ref{solfc}) by choosing the scalar functions in the form \cite{bb1}
\begin{eqnarray}\label{bes}
\chi_{q_z q_\perp m}^{\sigma}(\rho,\phi,z,t)
 = \rme^{-\rmi\sigma(\omega_qt- q_z z - m\phi)}J_{m}(q_\perp\rho),
\end{eqnarray}
where $\omega_q=cq=c\sqrt{q_\perp^2 + q_z^2}$ and $\sigma=\pm 1$ determines whether the wave will have the right-handed or the left-handed polarization. Since Bessel beams form a complete set of solutions of the Maxwell equations, every solution can be represented as a superposition of Bessel beams. Owing to the linear dependence of the RS vector on $\chi$, the construction of such superpositions can be realized by simply superposing the corresponding functions $\chi$. In order to obtain a wave packet of Bessel beams with a fixed angular momentum, we must include in this superposition only the Bessel functions with a given value of $m$. The general wave packet of this kind (for a fixed value of $\sigma$) is
\begin{eqnarray}\label{genbeam}
\fl \hspace{1cm}\chi_{m\sigma}(\rho,\phi,z,t)
= \rme^{\rmi\sigma m\phi}\int_0^\infty\!\!dq_\perp\int_{-\infty}^\infty\!\!dq_z\,h(q_z,q_\perp)
\rme^{-\rmi\sigma(c\sqrt{q_z^2+q_\perp^2} t - q_z z)}J_m(q_\perp\rho),
\end{eqnarray}
where $h(q_z,q_\perp)$ is some weight function. For beam-like solutions, the longitudinal component $q_z$ should not be spread out too much and also it must be much larger than the transverse component $q_\perp$.

There are very few weight functions that allow for an analytic evaluation of the integrals over $q_\perp$ and $q_z$ in (\ref{genbeam}). In addition to those treated in our previous work \cite{bb1}, there are several weight functions that lead to exponential beams. The simplest example of a beam with an exponential fall-off will be obtained by fixing the value of $q_z$ and taking a superposition of Bessel beams with a varying $q_\perp$. We shall only write down the formulas for $\sigma=1$. The case $\sigma=-1$ can be obtained by the complex conjugation. The expression (\ref{genbeam}) becomes now
\begin{eqnarray}\label{expbeam0}
\chi_{m q_z}(\rho,\phi,z,t) = \rme^{\rmi(m\phi+q_z z)}\!\int_0^\infty\!\!\!dq_\perp\,h(q_\perp)
\rme^{-\rmi c\sqrt{q_z^2+q_\perp^2} t}\!J_m(q_\perp\rho).
\end{eqnarray}
Obviously, the beam constructed in this way will not be monochromatic --- the weight function $h(q_\perp)$ determines the spectral characteristic of the beam. In order to show explicitly the distribution of frequencies, we change the integration variable from $q_\perp$ to $\omega=c\sqrt{q_z^2+q_\perp^2}$. In addition, for mathematical convenience, we shall replace the frequency $\omega$ by a dimensionless parameter $w=\omega/c\vert q_z\vert=\sqrt{1+(q_\perp/q_z)^2}$ and we rewrite (\ref{expbeam0}) in the form
\begin{eqnarray}\label{expbeam1}
\fl \hspace{1.5cm}\chi_{m q_z}(\rho,\phi,z,t) = \rme^{\rmi(m\phi+q_z z)}\int_{1}^\infty\!\!dw\,g(w) \rme^{-\rmi c\vert q_z\vert w t}J_m\left(\vert q_z\vert\rho\sqrt{w^2-1}\right),
\end{eqnarray}
where $g(w)$ is the spectral function. There are several spectral functions for which this integral can be evaluated analytically. The simplest example is (other functions are given in \ref{aa})
\begin{eqnarray}\label{expbeam2}
g(w)=\left(w^2-1\right)^{m/2}\!\!\rme^{-c\vert q_z\vert\tau w}.
\end{eqnarray}
\begin{figure}
\centering
\includegraphics[width=\textwidth]{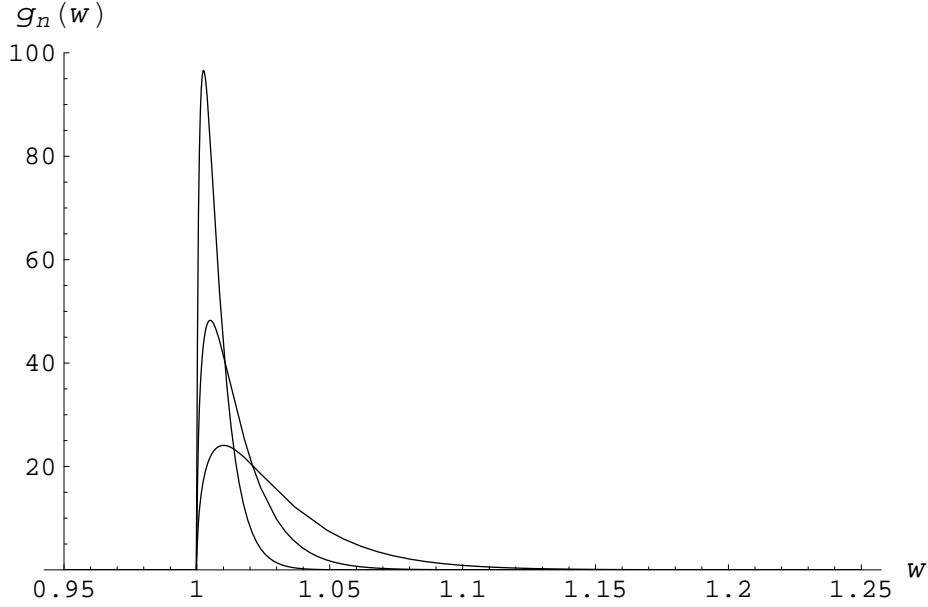}
\caption{The normalized spectral weights $g_n(w)$ plotted as functions of $w$ for $c q_z\tau=50$ (the lowest peak), $c q_z\tau=100$, and $c q_z\tau=200$.}\label{spectral}
\end{figure}
The parameter $\tau$ has the dimension of time and it determines the spread of frequencies $\Delta\omega=1/\tau$ in the wave packet. In figure \ref{spectral} we show the normalized spectral functions
\begin{eqnarray}\label{mac0}
\fl g_n(w)=\frac{\sqrt{\pi}(c \vert q_z\vert\tau/2)^{m/2+1/2}g(w)}{\Gamma(m/2+1)K_{m/2+1/2}(c \vert q_z\vert\tau)},\qquad
\int_1^\infty\!dw\,g_n(w)=1
\end{eqnarray}
for the three values of $\vert q_z\vert c\tau$. In order to better show the details, we have chosen in figure \ref{spectral} broadband beams (small $\tau$). Most beams used in realistic experiments are quasi monochromatic. For such beams, the characteristic dimensionless parameter $\vert q_z\vert c\tau$ is several orders of magnitude larger. The corresponding spectral functions would have the form of very high and very sharp needles. The integration over $w$ in (\ref{expbeam1}), for $g(w)$ given by (\ref{expbeam2}), can be performed due to the relation (cf. Eq.~6.645.2 of Ref.~\cite{gr})
\begin{eqnarray}\label{gr0}
\fl \int_1^\infty\!\!dx\,(x^2-1)^{\nu/2} \rme^{-\alpha x} J_\nu\left(\beta\sqrt{x^2-1}\right)
=\sqrt{\frac{2}{\pi}}\;
\frac{\beta^\nu K_{\nu+1/2}\left(\sqrt{\alpha^2+\beta^2}\right)}{(\alpha^2+\beta^2)^{\nu/2+1/2}}
\end{eqnarray}
and it leads to the following result for $\chi_{m q_z}$
\begin{eqnarray}\label{expbeam3}
\chi_{m q_z}(\rho,\phi,z,t) =\rme^{\rmi(m\phi+q_z z)}\sqrt{\frac{2}{\pi\vert q_z\vert}}
\frac{\rho^m K_{m+1/2}\left(\vert q_z\vert s\right)}{s^{m+1/2}},
\end{eqnarray}
where
\begin{eqnarray}\label{defs}
s=\sqrt{\rho^2 - c^2(t-\rmi\tau)^2}.
\end{eqnarray}
The Macdonald functions $K_{m+1/2}(x)$ for integer $m$ reduce to ordinary exponentials multiplied by polynomials in the inverse powers of the argument
\begin{eqnarray}\label{mac}
K_{m+1/2}(x)=\sqrt{\frac{\pi}{2x}}\rme^{-x}\sum_{k=0}^m\frac{(m+k)!}{k!(m-k)!}\frac{1}{(2 x)^k}.
\end{eqnarray}
Thus, the functions $\chi_{m q_z}$ describe beams with an exponential fall-off. Indeed, for large $\vert q_z\vert\rho$, but at the fixed values of all the remaining variables, these functions decrease exponentially
\begin{eqnarray}\label{expfall}
\chi_{m q_z}(\rho,\phi,z,t)\mathop{\sim}_{\rho\to\infty}
\rme^{\rmi(m\phi+q_z z)}\frac{\rme^{-\vert q_z\vert\rho}}{\vert q_z\vert\rho}.
\end{eqnarray}
The first three functions $\chi_{m q_z}$ (for $m=0,1,2$) are
\numparts\label{three}
\begin{eqnarray}
\chi_{0 q_z}(\rho,\phi,z,t) = \rme^{\rmi q_z z}\frac{\rme^{-\vert q_z\vert s}}{\vert q_z\vert s},\\
\chi_{1 q_z}(\rho,\phi,z,t) = \rme^{\rmi(\phi+q_z z)}\frac{\rho \rme^{-\vert q_z\vert s}}{(\vert q_z\vert s)^2}\left(1+\frac{1}{\vert q_z\vert s}\right),\\
\chi_{2 q_z}(\rho,\phi,z,t) = \rme^{\rmi(2\phi+q_z z)}\frac{\rho^2 \rme^{-\vert q_z\vert s}}{(\vert q_z\vert s)^3}\left(1+\frac{3}{\vert q_z\vert s}+\frac{3}{(\vert q_z\vert s)^2}\right).
\end{eqnarray}
\endnumparts
In the simplest case, when $m=0$, the RS vector is
\begin{eqnarray}\label{simpf}
\fl \left(\begin{array}{c}
F_\rho \\
F_\phi \\
F_z \end{array}\right)
=
\left(\begin{array}{c}
-\rmi q_z\rho(s^2+\vert q_z\vert s^3)\\
\rmi c(t-\rmi\tau)\rho(3+3\vert q_z\vert s+(\vert q_z\vert s)^2)\\
2(s^2+\vert q_z\vert s^3)-\rho^2(3+3\vert q_z\vert s+(\vert q_z\vert s)^2)
\end{array}\right)\frac{\rme^{\rmi q_z z}\rme^{-\vert q_z\vert s}}{\vert q_z\vert s^5}.
\end{eqnarray}
For nearly monochromatic exponential beams, i.e. for very large values of the parameter $\tau$, as compared with $\rho/c$ and $t$, the functions $\chi_{m q_z}$ have the form
\begin{eqnarray}\label{nearz}
\chi_{m q_z}(x,y,z,t)\mathop{\sim}_{\tau\to\infty}\frac{\rme^{-\vert q_z\vert c\tau}}{\vert q_z\vert (c\tau)^{m+1}}(x+\rmi y)^m\rme^{-\vert q_z\vert ct+\rmi q_z z},
\end{eqnarray}
clearly indicating a straight vortex line of strength $m$ along the $z$ axis. This is a general characteristic of all beams of radiation endowed with angular momentum \cite{bb1}. For $m=0$, we obtain the standard plane wave.

Having at our disposal the expansion (\ref{expbeam0}) of the exponential beam into the Bessel beams, we may calculate the spatial Fourier representation of the exponential beam. To this end, we use the expansion of the plane wave into Bessel beams (cf. Eq. 8.511.4 of Ref.~\cite{gr})
\begin{eqnarray}\label{expan}
\rme^{-\rmi{\bi k}\cdot{\bi r}}=\rme^{-\rmi k_z z}\sum_{m=-\infty}^{\infty}(-\rmi)^m\rme^{\rmi m(\varphi-\phi)}J_m(k_\perp\rho)
\end{eqnarray}
in the formula for the Fourier transform of $\chi_{m q_z}$
\begin{eqnarray}\label{fourier}
{\tilde \chi}_{m q_z}(k_\perp,\varphi,k_z,t)=\int\!d^3r\rme^{-\rmi{\bi k}\cdot{\bi r}}\chi_{m q_z}(\rho,\phi,z,t).
\end{eqnarray}
Upon substituting $\chi_{m q_z}$, as defined by (\ref{expbeam1}), into the formula (\ref{fourier}), we can easily perform the integrations over $\varphi$ and $z$ leading to delta functions. The remaining integration over $\rho$ produces also a delta function since
\begin{eqnarray}\label{deltab}
\int_0^\infty\!\rho\,d\rho J_m(k_\perp\rho)J_m(q_\perp\rho) = \frac{\delta(k_\perp-q_\perp)}{k_\perp}
\end{eqnarray}
and the final formula reads
\begin{eqnarray}\label{fourier1}
\fl {\tilde \chi}_{m q_z}(k_\perp,\varphi,k_z,t)
=(-\rmi)^m(2\pi)^2\delta(k_z-q_z)\frac{k_\perp^m}{\vert k_z\vert^{m+1}\sqrt{k_\perp^2+k_z^2}}\,
\rme^{\rmi m\varphi}\rme^{-c\sqrt{k_\perp^2+k_z^2}(\tau +\rmi t)}.
\end{eqnarray}
This result can also be obtained directly, without the use of the expansion (\ref{expbeam1}), by evaluating the Fourier transform of the original expression (\ref{expbeam3}) for $\chi_{m q_z}$, with the help of the formula 6.596.7 of Ref.~\cite{gr}.

\section{Time evolution of exponential beams}
\begin{figure}
\centering
\includegraphics[width=0.7\textwidth]{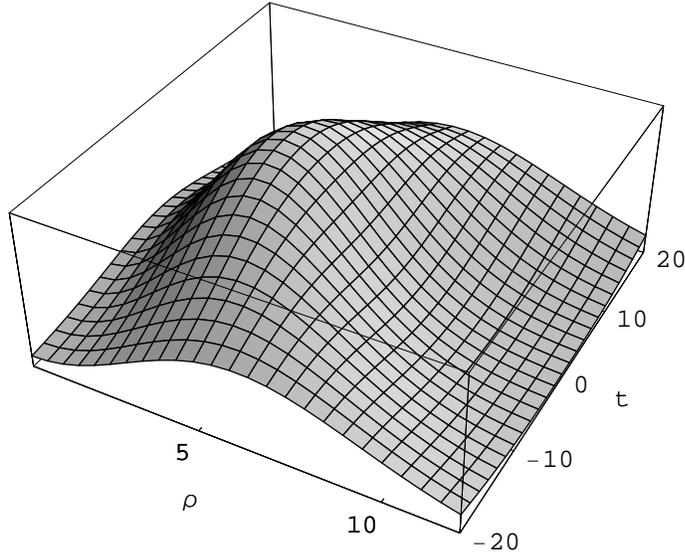}
\caption{The energy density ${\bm F}^*\cdot{\bm F}$ (in arbitrary units) shown as a function of the distance $\rho$ from the vortex center and time $t$.}\label{energy}
\end{figure}

\begin{figure}
\centering
\includegraphics[width=0.7\textwidth]{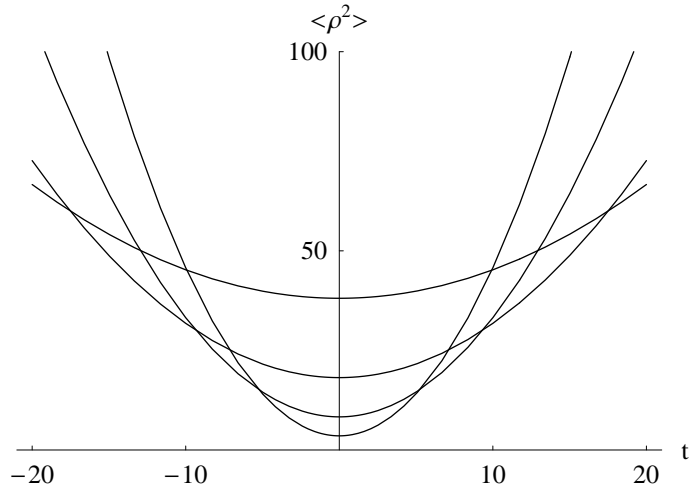}
\caption{Time dependence of the mean square radius for $m=1$ plotted for four values of the parameter $\tau=5,10,20,40$. Time and $\tau$ are measured in units of $1/c q_z$ and the radius is measured in units of $1/q_z$. The larger the value of $\tau$, the wider the parabola.}\label{radius}
\end{figure}

Exponential beams are not stationary --- they evolve in time. At a {\em given time}, our exponential beams look like diffraction-less beams --- their energy density and the Poynting vector {\em do not change} along the direction of propagation. In this respect they resemble Bessel beams. However, they exhibit a specific time dependence. In the transverse directions these beams first shrink and then expand. In our formulas the origin of time is chosen at the moment when the beam is maximally squeezed. This behavior is illustrated in figure \ref{energy} where we plot the energy density ${\bm F}^*\cdot{\bm F}$ calculated for the RS vector (\ref{simpf}) as a function of $\rho$ and $t$. In order to describe the time evolution in quantitative terms we shall use as a measure of the transverse size of the beam the mean square radius of the energy distribution
\begin{eqnarray}\label{tren1}
\langle x^2+y^2\rangle(t)=\int\!\int\!dxdy (x^2+y^2)\vert{\bm F}\vert^2/{\cal E}_{\rm tot}.
\end{eqnarray}
We prove in \ref{bb} that the time dependence of this quantity is simply quadratic, so that the formula for the time dependence reads
\begin{eqnarray}\label{tren}
\langle x^2+y^2\rangle(t)=\langle x^2+y^2\rangle(t=0)+2t^2{\cal E}_z/{\cal E}_{\rm tot},
\end{eqnarray}
where
${\cal E}_z$ is the energy carried by the $z$ components of the electromagnetic field and ${\cal E}_{\rm tot}$ is the total energy, both evaluated per unit length in the $z$ direction,
\begin{eqnarray}\label{full}
{\cal E}_z=\int\!\int\!dxdy \vert F_z\vert^2,\;\;\;{\cal E}_{\rm tot}=\int\!\int\!dxdy\vert{\bm F}\vert^2.
\end{eqnarray}
Several plots exhibiting the dependence on the parameter $\tau$ are shown in figure \ref{radius}.
Note that the more the beam is squeezed at $t=0$, the faster it expands.

\section{Conclusions}

We have constructed a new class of exact beam-like solutions of the Maxwell equations with an exponential decrease in the transverse direction. They are characterized by the wave vector $q_z$ and the projection of the angular momentum $m$ on the beam direction. These beams spread out in time but for a fixed value of $t$ the energy distribution does not change along the beam direction. The exponential beams are not monochromatic, but they can be made nearly monochromatic by taking a large value of the parameter $\tau$.

\ack
This research has been partly supported by the Polish Ministry of Scientific Research Grant Quantum Information and Quantum Engineering.

\appendix
\section{Further examples of exponential beams}\label{aa}

In addition to the simplest case described before, there are several other examples of beams with an exponential fall-off in the transverse direction. They are derived from the following formulas involving integrals of Bessel functions (cf. 6.645.1 and 6.646.1 of \cite{gr})
\numparts
\begin{eqnarray}
\fl \int_1^\infty\!\!dx\,\left(\frac{x-1}{x+1}\right)^{\!\nu/2}\!\!\!\!\rme^{-\alpha x} J_\nu\left(\beta\sqrt{x^2-1}\right)=\frac{\exp(-\sqrt{\alpha^2+\beta^2})}{\sqrt{\alpha^2+\beta^2}}
\left(\frac{\beta}{\sqrt{\alpha^2+\beta^2}+\alpha}\right)^{\!\nu}\!\!,\\\label{appa}
\int_1^\infty\!\!dx\,(x^2-1)^{-1/2} \rme^{-\alpha x} J_\nu\left(\beta\sqrt{x^2-1}\right)\nonumber\\
=I_{\nu/2}\left(\frac{\sqrt{\alpha^2+\beta^2}-\alpha}{2}\right)
K_{\nu/2}\left(\frac{\sqrt{\alpha^2+\beta^2}+\alpha}{2}\right).\label{appb}
\end{eqnarray}
\endnumparts
Upon choosing $g(w)$ in the form
\begin{eqnarray}\label{app2}
g(w)=\left(\frac{w-1}{w+1}\right)^{m/2}\!\!\rme^{-c\vert q_z\vert\tau w},
\end{eqnarray}
we obtain the following expression for an exponential beam
\begin{eqnarray}\label{app3}
\fl \chi_{m q_z}(\rho,\phi,z,t) =\frac{\rme^{\rmi(m\phi+q_z z)}\rme^{-\vert q_z\vert s}\rho^m}{\vert q_z\vert s\left(s-c(\tau+\rmi t)\right)^m}
=\frac{\rme^{\rmi(m\phi+q_z z)}\rme^{-\vert q_z\vert s}}{\vert q_z\vert s}\left(\frac{s+c(\tau+\rmi t)}{\rho}\right)^{\!m},
\end{eqnarray}
where $s$ was defined in (\ref{defs}).

The formula (\ref{appb}) requires several additional steps before leading to exponential beams. First, we shall put $\nu=1$ and express the Bessel functions $I_{1/2}$ and $K_{1/2}$ by elementary functions. This leads to
\begin{eqnarray}\label{app4}
\fl \int_1^\infty\!\!dx\,(x^2-1)^{-1/2} \rme^{-\alpha(x-1)} J_1\left(\beta\sqrt{x^2-1}\right)
=\frac{1}{\beta}\left(1-\rme^{\alpha-\sqrt{\alpha^2+\beta^2}}\right).
\end{eqnarray}
By differentiating this equation with respect to $\alpha$ and $\beta$ we may generate a whole family of new formulas. In particular, we obtain the formula
\begin{eqnarray}\label{app5}
\int_1^\infty\!\!dx\,(x^2-1)^{m/2}(x+1)^{-1} \rme^{-\alpha x} J_m\left(\beta\sqrt{x^2-1}\right)\nonumber\\
=-(-\beta)^m\left(\frac{d}{\beta d\beta}\right)^{\!m-1}\frac{\rme^{-\sqrt{\alpha^2+\beta^2}}}{\beta^2}
\left(1-\frac{\alpha}{\sqrt{\alpha^2+\beta^2}}\right).
\end{eqnarray}
that corresponds to
\begin{eqnarray}\label{app6}
g(w)=(w^2-1)^{m/2}(w+1)^{-1}\rme^{-c\vert q_z\vert\tau w},
\end{eqnarray}
and it leads to another exponential beam. General characteristics and the time evolution of all exponential beams are very similar. The overall behavior, as discussed in the main text, is determined by the exponential factor $\exp(-\vert q_z\vert s)$.

\section{Time dependence of the mean square radius}\label{bb}

We can find the time dependence of the transverse mean square radius by differentiating twice the expression
\begin{eqnarray}\label{defrsq}
\int\!\int\!dxdy (x^2+y^2)\vert{\bm F}\vert^2
\end{eqnarray}
with respect to time and using the Maxwell equations. This is done in three steps. First, we use the formulas
\begin{eqnarray}\label{step1}
\partial^2_t\vert{\bm F}\vert^2 &=& \partial^2_t{\bm F}^*\!\cdot\!{\bm F}+2\partial_t{\bm F}^*\!\cdot\!\partial_t{\bm F}+{\bm F}^*\!\cdot\!\partial^2_t{\bm F}\nonumber\\
&=& c^2\left(\Delta{\bm F}^*\!\cdot\!{\bm F}+2(\nabla\times{\bm F}^*)\!\cdot\!(\nabla\times{\bm F})+{\bm F}^*\!\cdot\!\Delta{\bm F}\right)\nonumber\\
&=& c^2\left(\Delta({\bm F}^*\!\cdot\!{\bm F}) - 2\partial_i\partial_j(F^*_iF_j)\right).
\end{eqnarray}
Note that the derivatives with respect to $\phi$ and $z$ drop out in the last line, because of the phase factor $\rme^{\rmi(q_z z+m\phi)}$ in ${\bm F}$. Next, we multiply the last line by $x^2+y^2$ and integrate over $x$ and $y$, to obtain after integration by parts
\begin{eqnarray}\label{step2}
\partial^2_t\int\!\int\!dxdy (x^2+y^2)\vert{\bm F}\vert^2 = 4\int\!\int\!dxdy\vert F_z\vert^2.
\end{eqnarray}
The last step is to prove that the last integral does not depend on time. To show this, we again use the Maxwell equations, integrate by parts, and add and subtract terms with derivatives with respect to $z$
\begin{eqnarray}\label{step3}
\fl \partial_t\int\!\int\!dxdy\vert F_z\vert^2 = \rmi c\int\!\int\!dxdy
\left(F_z^*(\partial_xF_y-\partial_yF_x)-(\partial_xF^*_y-\partial_yF^*_x)F_z\right)\nonumber\\
= \rmi c\int\!\int\!dxdy \left({\bm F}^*\!\cdot\!(\nabla\times{\bm F})+(F_y^*\partial_zF_x-F_x^*\partial_zF_y\right).
\end{eqnarray}
Since this expression is real, we may add its complex conjugate and take a half of the sum to obtain
\begin{eqnarray}\label{step4}
\fl \frac{\rmi c}{2}\int\!\int\!dxdy \left({\bm F}^*\!\cdot\!(\nabla\times{\bm F})-(\nabla\times{\bm F}^*)\!\cdot\!{\bm F})\right)
=-\frac{\rmi c}{2}\int\!\int\!dxdy \nabla\!\cdot\!({\bm F}^*\times{\bm F}) = 0.
\end{eqnarray}
As a byproduct of this calculation, we also obtain the time independence of ${\cal E}_{\rm tot}$ because the first integral (apart from 1/2) is the time derivative of ${\cal E}_{\rm tot}$. Putting it all together, we have
\begin{eqnarray}\label{step5}
\fl \int\!\int\!dxdy (x^2+y^2)\vert{\bm F}\vert^2=\left.\int\!\int\!dxdy (x^2+y^2)\vert{\bm F}\vert^2\right\vert_{t=0}+2c^2t^2\int\!\int\!dxdy\vert F_z\vert^2.
\end{eqnarray}
Dividing both sides of this equation by ${\cal E}_{\rm tot}$, and taking into account the time independence of ${\cal E}_{\rm tot}$, we obtain (\ref{tren}).

\end{document}